\documentclass[12pt]{article}
\usepackage{epsfig}
\newcommand{\spin}{\mathbf{S}}
\newcommand{\field}{\mathbf{h}}

\begin{document}

\title{\bf The stochastic traveling salesman problem:
           Finite size scaling and the cavity prediction}
\author{
Allon G.~Percus\\
{\normalsize CIC--3 and Center for Nonlinear Studies, MS-B258}\\
{\normalsize Los Alamos National Laboratory}\\
{\normalsize Los Alamos, NM 87545, USA}\\
{\normalsize E-mail: percus@lanl.gov}\\
\\
and \\
\\
Olivier C.~Martin\\
{\normalsize Division de Physique Th\'eorique}\\
{\normalsize Institut de Physique Nucl\'eaire, Universit\'e Paris-Sud}\\
{\normalsize F--91406 Orsay Cedex, France}\\
{\normalsize E-mail: martino@ipno.in2p3.fr}\\ }

\date{}
\maketitle
\begin{abstract}
\medskip
We study the random link traveling salesman problem, where lengths
$l_{ij}$ between city $i$ and city $j$ are taken to be independent,
identically distributed random variables.  We discuss a theoretical approach,
the cavity method, that has been proposed for finding the optimum tour
length over this random ensemble, given the assumption of replica
symmetry.  Using finite size scaling and a renormalized model, we test the
cavity predictions against the results of simulations, and find
excellent agreement over a range of distributions.  We thus provide numerical
evidence that the replica symmetric solution to this problem is the correct
one.  Finally, we note a surprising result concerning the
distribution of $k$th-nearest neighbor links in optimal tours, and invite
a theoretical understanding of this phenomenon.
 
\end{abstract} 

\bigskip
\par
\centerline{Key words: disordered systems, combinatorial optimization,
replica symmetry}
\vspace{1cm}
\centerline{Submitted to {\it Journal of Statistical Physics}, February 1998}
\centerline{Final version November 1998}
 
\newpage
\baselineskip=20pt

\section{Introduction}
\label{sec_introduction}

Over the past 15 years, the study of the traveling salesman problem (TSP)
from the point of view of statistical physics has been gaining added
currency, as theoreticians have improved their understanding of the relation
between combinatorial optimization and disordered systems.  The TSP may
be stated as follows: given $N$ sites (or ``cities''), find the total length
$L$ of
the shortest closed path (``tour'') passing through all cities exactly once.
In the stochastic TSP, the matrix of distances separating pairs of cities
is drawn randomly from an ensemble.  The ensemble that has received the
most attention in the physics community is the {\it random link\/} case,
where the individual lengths $l_{ij}$ between city $i$ and city $j$
($i<j$) are taken
to be independent random variables, all identically distributed according
to some $\rho(l)$.  The idea of looking at this random link ensemble,
rather than the more traditional ``random point'' ensemble where cities
are distributed uniformly in Euclidean space, originated with an attempt
by Kirkpatrick and Toulouse \cite{KirkpatrickToulouse} to find a version
of the TSP analogous to the earlier
Sherrington-Kirkpatrick (SK) model \cite{SherringtonKirkpatrick} for
spin glasses.

The great advantage of working with the random link TSP, rather than the
(random point) Euclidean TSP, is that one may realistically hope for
an analytical solution.  A major breakthrough occurred with the idea,
first formulated by M\'ezard and Parisi \cite{MezardParisi_86b} and later
developed by Krauth and M\'ezard \cite{KrauthMezard}, that the random link
TSP could be solved using the {\it cavity method\/}, an approach inspired
by work on spin glasses.  This method is based on assumptions pertaining
to properties of the system under certain limiting conditions.  The most
important of these assumptions is replica symmetry.
Although in the case of spin glasses, replica symmetry is violated
\cite{MezardParisiVirasoro}, for the TSP there are various grounds
for at least {\it suspecting\/} that replica symmetry
holds \cite{Sourlas,KrauthMezard}.
The cavity solution then leads to
a system of integral equations that can be solved --- numerically
at least --- to give a prediction of the optimum tour length $L$ in the
many-city limit $N\to\infty$.

In a previous article \cite{CBBMP}, we have taken
the random link distribution $\rho(l)$ to match that of the distribution
of individual city-to-city distances in the Euclidean case, 
using the random link TSP as a {\it random link approximation\/}
to the Euclidean TSP.  The approximation may seem crude
since it neglects all correlations between Euclidean distances, such
as the triangle inequality.  Nevertheless, it gives remarkably good
results.  In particular, a numerical solution of the 
random link cavity equations predicts
large $N$ optimum tour lengths that are within 2\% of the (simulated)
$d$-dimensional Euclidean values, for $d=2$ and $d=3$.
In the limit $d\to\infty$, this gap
shows all signs of disappearing.  The random link problem, and its
cavity prediction, is thus more closely related to the Euclidean problem
than one might expect.

The random link TSP is also, however, interesting in itself.
Little numerical work has accompanied the analytical
progress made --- a shortcoming made all the more troubling by the
uncertainties surrounding the cavity method's assumptions.
In this paper we attempt to redress the imbalance,
providing a numerical study of the finite size scaling of
the random link optimum tour length, and arguments
suggesting that the cavity solution is in fact correct.
In the process, our numerics reveal some remarkable properties concerning
the frequencies with which cities are connected to their $k$th-nearest
neighbor in optimal tours; we invite a theoretical explanation of these
properties.

\section{Background and the cavity method}
\label{sec_cavity}

In an attempt to apply tools from statistical mechanics to optimization
problems, Kirkpatrick and Toulouse \cite{KirkpatrickToulouse} introduced
a particularly
simple case of the random link TSP.  The distribution of link lengths
$l_{ij}$ was taken to be uniform, so that $\rho(l)$ is constant over
a fixed interval.  In light of the random link approximation, one may
think of this as corresponding, at large $N$, to the 1-D Euclidean case.
(When cities are randomly and uniformly distributed on a line segment,
the distribution of lengths between pairs of cities is uniform.)
Although the 1-D Euclidean case is trivial --- particularly
if we adopt periodic boundary conditions, in which case the optimum tour
length is simply the length of the line segment --- the corresponding
random link problem is far from trivial.

The simulations performed by Kirkpatrick and Toulouse
suggested a random link optimum tour
length value of $L_{RL}\approx 1.045$ in the $N\to\infty$ limit.\footnote{Here
we work in units where the line segment is taken to have unit length,
and in order to match the normalized 1-D Euclidean distribution, 
we let $\rho(l)=2$ on $[0,1/2)$.  The 1-D Euclidean value, for comparison,
would thus be $L_E=1$.  Kirkpatrick and Toulouse, among others, choose instead
$\rho(l)=1$ on $[0,1)$, contributing an additional factor of 2 in $L_{RL}$
which we omit when quoting their results.}
M\'ezard and Parisi \cite{MezardParisi_86a} attempted to improve both upon this
estimate and upon the theory by 
using replica techniques often employed in spin glass problems (for a discussion
of the replica method in this context, see \cite{MezardParisiVirasoro}).
This approach
allowed them to obtain, via a saddle point approximation, many orders of
the high-temperature expansion for the internal energy.  They then
extrapolated down to zero temperature --- corresponding to
the global TSP optimum --- finding $L_{RL}=1.04\pm 0.015$.  Their analysis,
like that of Kirkpatrick and Toulouse, was carried out only for the case
of $\rho(l)$ equal to a constant.

Given the difficulties of pushing the replica method further, M\'ezard
and Parisi then
tried a different but related approach known as the {\it cavity\/} method
\cite{MezardParisi_86b}.
This uses a mean-field approximation which, in the case of spin glasses,
gives the same result as the replica method in the thermodynamic limit
($N\to\infty$).  As much of the literature on the cavity method has been
prohibitively technical to non-specialists, we shall review
the approach in more conventional language here, indicating what is involved
in the case of the TSP.

Both the replica and the cavity methods involve a representation
of the partition function originally
developed in the context of polymer theory \cite{DeGennes,Orland}.  The
approach consists of mapping the TSP onto an $m$-component spin system,
writing
down the partition function at temperature $T$, and then taking the limit
$m\to 0$.  More explicitly, consider
$N$ spins $\spin_i$, $i=1,\dots,N$ (corresponding to the $N$ cities), where
each spin $\spin_i$ has $m$
components $S_i^\alpha$, $\alpha=1,\dots,m$, and where
$(\spin_i)^2=m$ for all $i$.  The partition
function is defined, in terms of a parameter $\omega$, as
\begin{eqnarray}
\label{eq_partfuncunexpanded}
Z &=& \int\prod_q d\mu(\spin_q )\,\exp(\omega\,\sum_{i<j}R_{ij}\,
\spin_i\cdot\spin_j)\\
&=& \int\prod_q d\mu(\spin_q )\left[1+\omega\,\sum_{i<j}R_{ij}
(\spin_i\cdot
\spin_j) + \frac{\omega^2}{2!}\,\sum_{\scriptstyle{i<j}\atop\scriptstyle{k<l}}
R_{ij}\,R_{kl}
(\spin_i\cdot\spin_j)(\spin_k\cdot\spin_l)+\cdots
\right]
\label{eq_partfunc}
\end{eqnarray}
where the integral is taken over all possible values spin values
(the area measure is normalized so that $\int d\mu(\spin_q ) = 1$),
and $R_{ij}$ is related to the length $l_{ij}$ between city
$i$ and city $j$ as $R_{ij}\equiv e^{-N^{1/d}l_{ij}/T}$.  Now
employ a classic diagrammatic argument: let each spin product
$(\spin_a\cdot\spin_b)$ appearing in the series be represented by an
edge in a graph whose vertices are the $N$ cities. The first-order
terms ($\omega$)
will consist of one-edge diagrams, the second-order terms ($\omega^2$)
will consist of two-edge diagrams, and so on.
What then happens when we integrate over all spin
configurations?  If there is a spin $\spin_a$ that
occurs only once in a given diagram, {\it i.e.\/}, it is an endpoint,
the spherical symmetry of $\spin_a$ will
cause the whole expression to vanish.  The non-vanishing summation terms in
(\ref{eq_partfunc}) therefore correspond only to ``closed'' diagrams,
where there is at least one loop.  It may furthermore
be shown that in performing the integration, any one of these
closed diagrams will contribute a factor $m$ for every loop present
in the diagram \cite{Orland}.  If we then consider $(Z-1)/m$ and take
the limit $m\to 0$,
it is clear that only diagrams with a single loop will remain.
Furthermore, since any closed diagram with more than $N$ links must
necessarily contain more than one loop, only diagrams up to order $\omega^N$
will remain.  Finally, take the limit $\omega\to\infty$.  The term that
will then dominate in (\ref{eq_partfunc}) is the order $\omega^N$ term
which, being a single loop diagram, represents precisely a closed tour
passing through all $N$ cities.  We may write it without the combinatorial
factor $N!$ by expressing it as a sum over ordered pairs in the tour,
and we thus find:
\begin{eqnarray}
\lim_{\scriptstyle{m\to 0}\atop\scriptstyle{\omega\to\infty}}
\frac{Z-1}{m\omega^N} &=&
\sum_{\scriptstyle{N\mbox{\scriptsize -link single loops}}\atop
\scriptstyle{(i_1,i_2,\dots,i_N)}}
R_{i_1i_2}\,R_{i_2i_3}\cdots R_{i_{N-1}i_N}\,R_{i_Ni_1}\\
&=& \sum_{N\mbox{\scriptsize -city tours}} e^{-N^{1/d}L/T}
\label{eq_ztsp}
\end{eqnarray}
where $L$ is the total tour length.  What we obtain is exactly the partition
function for the traveling salesman problem, with the correct canonical
ensemble Boltzmann weights, using the tour length as the energy to be
minimized (up to a factor $N^{1/d}$, necessary for the energy to be
extensive).

The idea behind the cavity method is as follows.
Since all spin couplings $R_{ij}$ in
(\ref{eq_partfuncunexpanded}) are positive (ferromagnetic), we expect the
$m$-component spin system to have a non-zero spontaneous magnetization
in equilibrium.  Now add
an $(N+1)$th spin to the system; it too acquires a spontaneous magnetization
$\langle\spin_{N+1}\rangle$.  Let us obtain the thermodynamic
observables of the new system (in particular $\langle\spin_{N+1}\rangle$ itself)
in terms of the earlier magnetizations $\langle\spin_i\rangle'$
from {\it before} the $(N+1)$th spin was added --- hence the notion of
a ``cavity''.

In order to compute these
relations, an important mean-field assumption is made: that at large $N$,
any effect spin $N+1$ feels from correlations among the $N$ other spins
is negligible.  We justify this in the following way.
Although all spins in (\ref{eq_partfuncunexpanded}) are indeed coupled,
the coupling constants $R_{ij}$ decrease exponentially with length $l_{ij}$,
and so effective interactions arise only between very near neighbors.  But
a crucial property of the random link model is that the near neighbors
of spin $N+1$ are {\it not\/} generally near neighbors of one another:
they are near neighbors of one another only with probability
$O(1/N)$.  Thus, when considering quantities involving spin
$N+1$, the effect of direct interactions between any two of its neighbors
is $O(1/N)$, and decays to zero in the limit $N\to\infty$.
We therefore replace (\ref{eq_partfuncunexpanded}) by the mean-field partition
function
\begin{equation}
\label{eq_ZMF}
Z_{MF} = \int\prod_{q=1}^N d\mu(\spin_q )
\int d\mu(\spin_{N+1})\,\exp(\omega\,\sum_{i=1}^N R_{i,N+1}\,
\spin_i\cdot\spin_{N+1} + \sum_{i=1}^{N+1}\spin_i\cdot\field_i).
\end{equation}

By definition, if spin $N+1$ were removed, we would recover the ``cavity
magnetizations'' $\langle\spin_i\rangle'$.  This requirement is sufficient
to specify the fields $\field_i$.  Stripping
out spin $N+1$ from (\ref{eq_ZMF}) leaves us simply with a product
of integrals $\prod_q\int d\mu(\spin_q)\exp(\spin_q\cdot\field_q)$,
whose logarithmic derivative with respect to $\field_i$ must then give the
magnetization $\langle\spin_i\rangle'$.  We may obtain this expression
by expanding the integrands, taking advantage of the identity
$\int d\mu(\spin_i)\, S_i^{\alpha} S_i^{\beta}= \delta_{\alpha\beta}$
for all spin components $\alpha$ and $\beta$, as well as the nilpotency
property~\cite{MezardParisi_86b} that in the limit $m\to 0$, integrating
the product of {\it more\/} than two components of $\spin_i$ gives zero.
(This is analogous to the property used earlier in the
diagrammatic expansion.)  The result is
\begin{equation}
\langle\spin_i\rangle' = \frac{\field_i}{1+(\field_i)^2/2}.  
\end{equation}
Note that this specifies $\field_i$ for $1\le i\le N$; $\field_{N+1}$
has been introduced purely for analytical convenience, and will ultimately
be set to 0.

Without loss of generality, let us assume the spontaneous magnetizations
of the system to be directed exclusively along component 1.
This may be imposed, for instance, by applying an additional infinitesimal
field directed along component 1.  Physically, however, the assumption that
distant spins are uncorrelated also means that this infinitesimal field is
sufficient to select a single phase or equilibrium state, thus giving rise
to a unique thermodynamic limit.  From the point of view
of dynamics, a consequence is that two ``copies'' of the system will evolve
to the same equilibrium distribution.
This property is known as {\it replica symmetry\/}, and has been central to
the modern understanding of disordered systems.%
\footnote{An analogous property was used to obtain the
replica solution mentioned earlier.}
Replica symmetry is in fact known to be broken in spin glasses; if one uses,
for instance, the replica symmetric solution of the SK model, one finds
a ground state energy prediction
that is inaccurate by about $5\%$ \cite{MezardParisiVirasoro}.  However,
this does not mean that replica symmetry breaking
occurs in {\it all\/} related
problems of high complexity (the TSP and the spin glass both fall into the
{\it NP-hard\/} class of computational complexity).
Showing that the (replica symmetric) cavity solution correctly predicts
macroscopic
quantities for the random link TSP would suggest that the TSP, unlike a
spin glass, does indeed exhibit replica symmetry.

In order to obtain the cavity solution,
consider the mean-field expression (\ref{eq_ZMF}).
Taking advantage of nilpotency, as well as the fact that $R_{ij}$
is non-negligible only with probability $O(1/N)$, we may expand (\ref{eq_ZMF})
and obtain in the large $N$ limit:
\begin{eqnarray}
\label{eq_ZMFexpand}
Z_{MF}&=&\prod_{i=1}^N \left(1+\frac{(\field_i)^2}{2}\right) \\
&&\times\left[ 1\,+\,\frac{(\field_{N+1})^2}{2}\,+\,
\sum_{j=1}^N\frac{\omega R_{j,N+1}\,\field_j\cdot\field_{N+1}}
{1+(\field_j)^2/2} +
\sum_{1\le j<k\le N}\frac{\omega^2 R_{j,N+1} R_{k,N+1}\,\field_j\cdot\field_k}
{[1+(\field_j)^2/2][1+(\field_k)^2/2]}\right].\nonumber
\end{eqnarray}
Differentiating $Z_{MF}$ with respect to $\field_{N+1}$ and
then setting $\field_{N+1}=0$ yields an expression for
$\langle\spin_{N+1}\rangle$ in terms of the remaining $\field_i$, or
equivalently in terms of the cavity magnetizations $\langle\spin_i\rangle'$.
This expression simplifies further simplifies at large $\omega$.
Recalling that the magnetization is by construction directed along
component 1, we obtain~\cite{MezardParisi_86b}:
\begin{equation}
\label{eq_recur1}
\langle S_{N+1}^1\rangle = \frac{\sum_{j=1}^N 
R_{j,N+1}\langle S_j^1\rangle'}
{\omega\,\sum_{1\le j<k\le N}
R_{j,N+1} R_{k,N+1} \langle S_j^1\rangle'\,\langle S_k^1\rangle'}.
\end{equation}
(The factor $\omega$ in the denominator may be avoided, if need be, by
applying a uniform rescaling factor $\sqrt{\omega}$ to all magnetizations.)
Thus, using the mean-field approach, we can express the
magnetization of the $(N+1)$th spin in terms of what the other magnetizations
would be in the {\it absence\/} of this $(N+1)$th spin.

While these quantities have been derived for a spin system whose
partition function is given by $Z$, we are interested in the TSP
whose partition function is given by (\ref{eq_ztsp}).  Consider an important
macroscopic quantity for the TSP: the frequency with which a tour occupies
a given link.  Define $n_{ij}$ to be $1$ if the link ${ij}$ is
in the tour, and $0$ otherwise.  Since the total tour length (energy) is
$L = \sum_{i<j} n_{ij} l_{ij}$, the {\it mean\/} occupation
frequency $\langle n_{ij}\rangle$, averaged over all tours with the
Boltzmann factor, is simply found from the logarithmic derivative
of (\ref{eq_ztsp}) with respect to $l_{ij}$.  Using $Z_{MF}-1$ in
place of $Z-1$, and proceeding as above, we obtain in the limit
$\omega\to\infty$:
\begin{equation}
\label{eq_recur2}
\langle n_{i,N+1} \rangle = R_{i,N+1}
\langle S_i^1\rangle'
\frac{\sum_{j\ne i}R_{j,N+1}\langle S_j^1\rangle'}
{\sum_{1\le j<k\le N}
R_{j,N+1} R_{k,N+1} \langle S_j^1\rangle'\,\langle S_k^1\rangle'}
\mbox{.}
\end{equation}

The relations (\ref{eq_recur1}) and (\ref{eq_recur2})
have been derived for a single realization of the $R_{ij}$'s.
In the ensemble of instances we consider here, the 
thermal averages become random variables
with a particular distribution. 
As far as
(\ref{eq_recur1})
is concerned, we may treat the magnetizations $\langle S_i^1\rangle'$ as
independent identically distributed random variables.  Furthermore, the
existence of a thermodynamic limit in the model requires that at large
$N$, $\langle S_{N+1}^1\rangle$ have the {\it same\/} distribution as the
cavity magnetizations; this
imposes, for a given link length distribution $\rho(l)$, a unique
self-consistent probability distribution of the magnetizations.  From
(\ref{eq_recur2}), one can then find the probability distribution of
$\langle n_{N+1,i}\rangle$, and in turn, taking the $T\to 0$ limit, the
distribution
${\cal P}(l)$ of link lengths $l$ used {\it in the optimal tour\/} (at
$N\to\infty$).

Krauth and M\'ezard~\cite{KrauthMezard} carried
out this calculation,
for $\rho(l)$ corresponding
to that of the $d$-dimensional Euclidean case, namely
\begin{equation}
\label{eq_dist}
\rho_d(l)=\frac{2\,\pi^{d/2}}{\Gamma(d/2)}\,l^{d-1}\mbox{.}
\end{equation}
Of course, $\rho_d(l)$ must be cut off at some finite $l$ in
order to be normalizable; precisely how this is done is unimportant,
however, since only the behavior of $\rho_d(l)$ at small $l$ is relevant
for the optimal tour in the $N\to\infty$ limit.
The result of Krauth and M\'ezard's calculation is:
\begin{eqnarray}
{\cal P}_d(l)&=&N^{-1/d}\pi^{d/2}\,\frac{\Gamma(d/2+1)}{\Gamma(d+1)}\,
\frac{l^{d-1}}{2\Gamma(d)}\nonumber \\
&&\times\left(-\frac{\partial}{\partial l}\right)
\int_{-\infty}^{+\infty}\left[ 1+H_d(x)\right] e^{-H_d(x)}
\left[ 1+H_d(l-x))\right] e^{-H_d(l-x)}\,dx\mbox{,}
\label{eq_lldist}
\end{eqnarray}
where $H_d(x)$ is the solution to the integral equation
\begin{equation}
H_d(x)=\pi^{d/2}\,\frac{\Gamma(d/2+1)}{\Gamma(d+1)}
\int_{-x}^{+\infty}\frac{(x+y)^{d-1}}{\Gamma(d)}\,
\left[ 1+H_d(y)\right]\, e^{-H_d(y)}\,dy\mbox{.}
\end{equation}
From ${\cal P}_d(l)$, one may obtain the mean link length in the tour,
and thus the cavity prediction $L_{RL}^c$ for the total length of the tour.
Introducing the large $N$ asymptotic quantity 
$\beta_{RL}(d)\equiv\lim_{N\to\infty}L_{RL}(N,d)/N^{1-1/d}$, the cavity
prediction $\beta_{RL}^c(d)$ is then:
\begin{eqnarray}
\beta_{RL}^c(d)&=&\lim_{N\to\infty}\, N^{1/d}\int_0^{+\infty} l\,{\cal P}_d(l)\,dl\nonumber \\
&=&\frac{d}{2}\,\int_{-\infty}^{+\infty}
H_d(x)\left[ 1+H_d(x)\right] e^{-H_d(x)}\,dx\mbox{.}
\end{eqnarray}

At $d=1$, Krauth and M\'ezard solved these equations numerically,
obtaining $\beta_{RL}^c(1) = 1.0208\ldots$  It is difficult to compare this
with Kirkpatrick's value of $\beta_{RL}(1)\approx 1.045$ from direct
simulations (as no error estimate exists for the latter quantity), however
an analysis \cite{Percus_Thesis} of recent numerical results by Johnson
{\it et al.\/} \cite{Johnson_HK} gives $\beta_{RL}(1)=1.0209\pm 0.0002$,
lending strong credence to the cavity value.
Krauth and M\'ezard also performed
a numerical study of ${\cal P}_1(l)$.  They found the
cavity predictions to be in good agreement
with what they found in their own direct simulations.  Further numerical
evidence supporting the assumption of replica symmetry was found by
Sourlas \cite{Sourlas}, in an investigation of the low temperature statistical
mechanics of the system.  Thus, for the $l_{ij}$ distribution at
$d=1$, there is good reason to believe that the cavity
assumptions are valid and that the resulting predictions are exact
at large $N$, so that $\beta_{RL}^c(1)=\beta_{RL}(1)$.

At higher dimensions, the values of $\beta_{RL}^c(d)$
were given by the present
authors in \cite{PercusMartin_PRL}, and a large $d$ power series solution
for $\beta_{RL}^c(d)$ was derived \cite{Boutet_Thesis,CBBMP}:
\begin{equation}
\label{eq_cavity}
\beta_{RL}^c(d)=\sqrt{\frac{d}{2\pi e}}\,(\pi d)^{1/2d}\,
\left[1+\frac{2-\ln 2-2\gamma}{d}+O\left(\frac{1}{d^2}\right)\right]\mbox{,}
\end{equation}
where $\gamma$ represents Euler's constant ($\gamma = 0.57722\ldots$).
But is the cavity method exact --- that is, is
$\beta_{RL}^c(d)=\beta_{RL}(d)$ --- for {\it all\/} $d$,
or is $d=1$ simply a pathological case (as it is in the Euclidean model,
where $\beta_E(1)=1$)?
While it appears sensible to argue that the qualitative properties of
the random link TSP are insensitive to $d$, there is as yet no evidence
that replica symmetry holds for $d \ne 1$.  Our purpose here is to
provide such evidence by numerical simulation, as has been done,
for instance, in a
related combinatorial optimization problem known as the matching
problem \cite{BKMP,Boutet_Thesis}.
We now turn to this task, considering first the $d=2$ case, and then a
``renormalized'' random link model that enables us to verify numerically
the $O(1/d)$ coefficient predicted in (\ref{eq_cavity}).

\section{Numerical analysis: $d=2$ case}
\label{sec_numerd2}

We have implicitly been making the assumption so far, via our notation,
that as $N\to\infty$ the random variable $L_{RL}(N,d)/N^{1-1/d}$
approaches a unique value $\beta_{RL}(d)$ with probability 1.
This is a property known as self-averaging.  The analogous
property has been shown for the Euclidean TSP at all
dimensions \cite{BHH}.  For the random link TSP, however, the only
case where a proof of self-averaging is known is in the $d\to\infty$ limit,
where a converging upper and lower bound give in fact the {\it exact\/}
result \cite{VannimenusMezard}:
\begin{equation}
\label{eq_rlexact}
\beta_{RL}(d)=\sqrt{\frac{d}{2\pi e}}\,(\pi d)^{1/2d}
\left[ 1 + O\left( \frac{1}{d}\right) \right]\mbox{.}
\end{equation}
Comparing this with (\ref{eq_cavity}), we may already see
that $\beta_{RL}^c(d)\sim\beta_{RL}(d)$ when $d\to\infty$, and
so the cavity prediction is correct in the infinite dimensional limit.

For finite $d$, however, it has not been shown analytically that
$\beta_{RL}(d)$
even exists.  To some extent, the difficulty in proving this can be
traced to the non-satisfaction of the triangle inequality.  The reader
acquainted with the self-averaging proof for the Euclidean TSP
may see that the ideas
used there are not applicable to the random link case; for instance,
combining good subtours using simple insertions will not lead to
near-optimal global tours, making the problem particularly challenging.
Let us therefore examine the distribution of $d=2$ optimum tour lengths
using numerical simulations, in order to give empirical support for the
assertion that the $N\to\infty$ limit is well-defined.

\begin{figure}[t]
\begin{center}
\epsfig{file=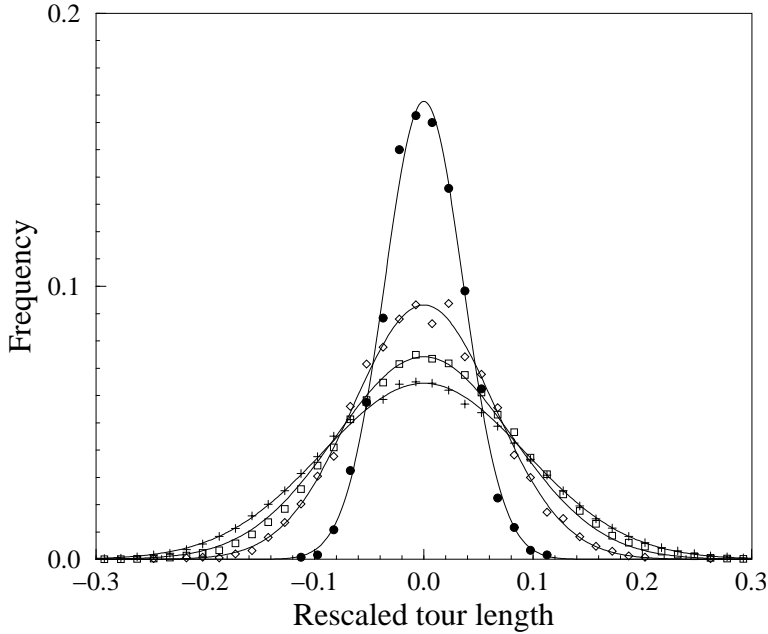,width=4in}
\caption{\small Distribution of 2-D random link rescaled tour length
$(L_{RL}-\langle
L_{RL}\rangle)/\sqrt{N}$ for increasing values of N.  Plus signs show
$N=12$ (100,000 instances used), squares show $N=17$ (100,000 instances
used), diamonds show $N=30$ (4,000 instances used), and dots show $N=100$
(1,200 instances used).
Solid lines represent Gaussian fits for each value of $N$ plotted.}
\label{fig_rl_selfavd2}
\end{center}
\end{figure}

The algorithmic procedures we use for simulations are identical to those
we have used in an earlier study concerning the Euclidean TSP \cite{CBBMP};
for details, the interested reader is referred to that article.  Briefly
stated, our optimization procedure involves using
the LK and CLO local search heuristic algorithms
\cite{LinKernighan,MartinOtto_AOR} where for each instance of the
ensemble we run the
heuristic over multiple random starts.  LK is used for smaller values
of $N$ ($N\le 17$) and CLO, a more sophisticated method combining LK 
optimization with random jumps, for larger values of $N$ ($N=30$ and
$N=100$).
There is, of course, a
certain probability that even over the course of multiple random starts,
our heuristics will not find the true optimum of an instance.  We
estimate the associated systematic bias using a number of test instances,
and adjust the number of random starts to keep this bias at least an
order of magnitude below other sources of error discussed below.
(At its maximum --- occurring in the $N=100$ case --- the systematic
bias is estimated as under 1 part in 20,000.)

\begin{table}[!b]
\vspace{.5in}
\caption{\small Variance of the non-rescaled optimum tour length $L_{RL}(N,2)$
with increasing $N$.}
\label{table_rl_vard2}
\begin{center}
\begin{tabular}{ccc}
$N$&$\sigma^2$&\# instances used\\
\hline
12&$0.3200$&$100$,$000$\\
17&$0.3578$&$100$,$000$\\
30&$0.3492$&$4$,$000$\\
100&$0.3490$&$1$,$200$\\
\end{tabular}
\end{center}
\end{table}

Following this numerical method, we see from our simulations (Figure
\ref{fig_rl_selfavd2})
that the distribution of $L_{RL}(N,2)/\sqrt{N}$ becomes increasingly
sharply peaked for increasing $N$, so that the ratio approaches 
a well-defined limit
$\beta_{RL}(2)$.  Furthermore, the variance of $L_{RL}(N,2)$ remains relatively
constant in $N$ (see Table \ref{table_rl_vard2}), indicating that the
width $\sigma$ for the distribution shown in the figure decreases as
$1/\sqrt{N}$, strongly suggesting a Gaussian distribution.  Similar
results were found
in our Euclidean study
(albeit in that case with $\sigma$ being approximately half of its random
link value).  This is precisely the sort of behavior one would expect
were the central limit theorem to be applicable.

\begin{figure}[t]
\begin{center}
\epsfig{file=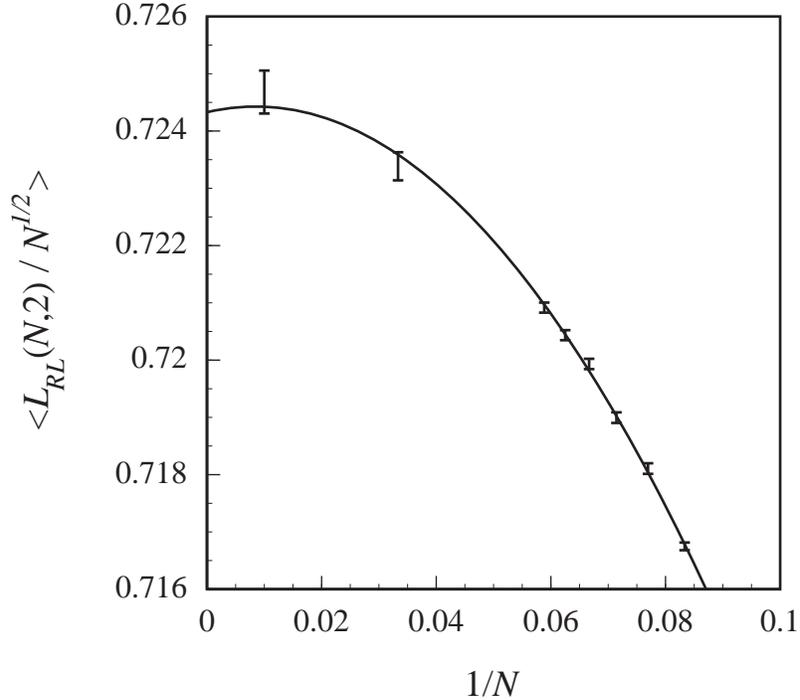}
\caption{\small Finite size scaling of mean optimum tour length for $d=2$.
Best fit ($\chi^2=4.46$)
is given by:
$\langle L_{RL}(N,2)\rangle/N^{1/2} = 0.7243
(1 + 0.0322/N - 1.886/N^2)$.  Error bars show one standard deviation
(statistical error).}
\label{fig_rl_fssd2}
\end{center}
\end{figure}

Let us now consider the large $N$ limit of $L_{RL}(N,2)/\sqrt{N}$, as given by
numerical simulations.  In the Euclidean case, it has been observed
\cite{CBBMP} that the
finite size scaling law can be written in terms of a power series in
$1/N$.  The same arguments given there apply to the random link case, and
so we may expect the ensemble average $\langle L_{RL}(N,2)\rangle$ to satisfy
\begin{equation}
\langle L_{RL}(N,d)\rangle=\beta_{RL}(d)\,N^{1-1/d}\,
\left[ 1+\frac{A(d)}{N} + \cdots
\right]\mbox{.}
\end{equation}
In order to obtain $\langle L_{RL}(N,2)\rangle$ at a finite value
of $N$ from simulations,
we average over a large number of instances to reduce the statistical
error arising from instance-to-instance fluctuations.  Figure
\ref{fig_rl_fssd2}
shows the results of this, with accompanying error bars, fitted to the
expected finite size scaling law (truncated after $O(1/N^2)$).  The fit
is a good one: $\chi^2=4.46$ for 5 degrees of freedom.  As in \cite{CBBMP},
we may obtain an error estimate on $\beta_{RL}(2)$ by noting that if we
take the extrapolated value and add or subtract one standard deviation, and
then redo the fit with this as a fixed constant, $\chi^2$ will increase by 1.
We thus find $\beta_{RL}(2)=0.7243\pm 0.0004$, in very good agreement
with the cavity result of $\beta_{RL}^c(2)=0.7251\ldots$
The discrepancy between the two is consistent with the statistical error (two
standard deviations apart), and in relative terms is approximately $0.1\%$.
The fit in Figure \ref{fig_rl_fssd2}, furthermore, appears robust with
respect to sub-samples of the data; even if we disallow the use of the
$N=100$ data point in the fit, the resulting asymptotic value is still within
$0.25\%$ of the cavity prediction.  By comparison, recall that the error in the
replica symmetric solution to the SK spin glass ground state energy is of the
order of $5\%$ \cite{MezardParisiVirasoro}.

\begin{figure}[!b]
\begin{center}
\epsfig{file=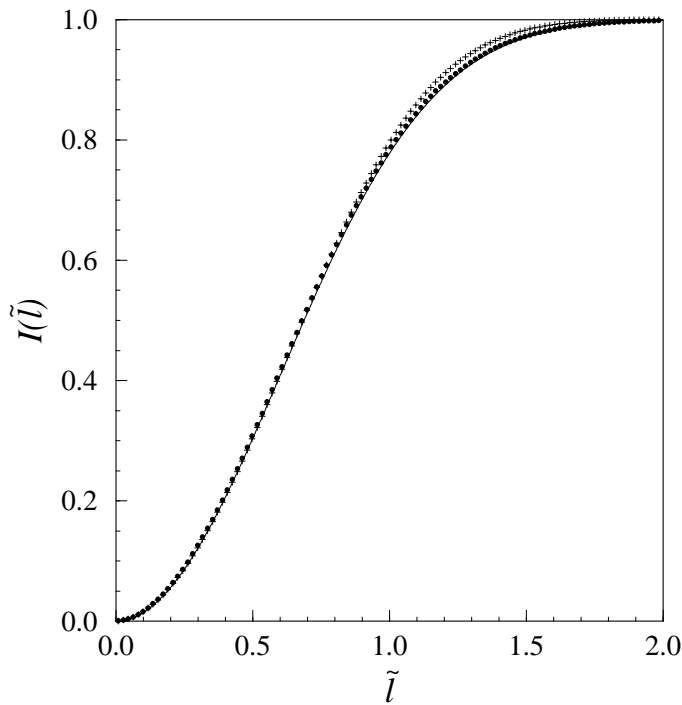,width=3.5in}
\caption{\small Integrated probability distribution of link lengths in the
optimal tour, for $d=2$, using rescaled length $\tilde{l}=l\sqrt{N}$.
Plus signs represent $N=12$ simulation results, dots represent $N=100$
simulation results, and solid line represents cavity prediction.}
\label{fig_rl_distscale}
\end{center}
\end{figure}

Another quantity that Krauth and M\'ezard studied in their $d=1$ numerical
investigation \cite{KrauthMezard} was the optimum tour link
length distribution
${\cal P}_d(l)$ given in (\ref{eq_lldist}).  Let us consider ${\cal P}_2(l)$,
and following their example, let us look specifically at the integrated
distribution $I_d(l)\equiv\int_0^l {\cal P}_d(l')\,dl'$.
The cavity
result for $I_d(l)$ can, like $\beta_{RL}(d)$, be computed numerically
to arbitrary precision.
In Figure \ref{fig_rl_distscale} we compare this with the results of direct
simulations, for $d=2$, at increasing values of $N$.
The improving agreement for
increasing $N$ (within 2\% at $N=100$) strongly suggests that the
cavity solution gives the exact $N\to\infty$ result.

\begin{figure}[t]
\begin{center}
\epsfig{file=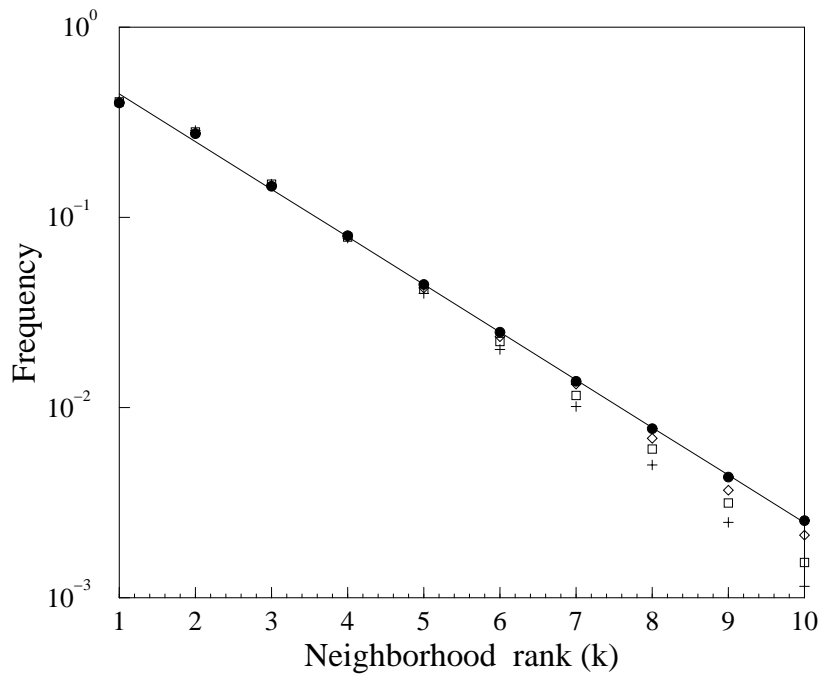,width=4in}
\caption{\small Frequencies with which $k$th-nearest neighbors are used in
optimal 2-D random link tours.  Plus signs show values for $N=12$,
squares for $N=17$, diamonds for $N=30$, and dots for $N=100$.  Best
exponential fit (straight line on log plot) is shown for $N=100$ data.}
\label{fig_rl_neighbord2}
\end{center}
\end{figure}

Finally, it is of interest to consider one further quantity in
the $d=2$ random link simulations, for which there is at present no
corresponding cavity prediction: the frequencies of ``neighborhood rank''
used in the optimal tour, that is, the proportion of links connecting
nearest neighbors, 2nd-nearest neighbors, etc.
Sourlas \cite{Sourlas} has noted
that in practice in the $d=1$ case, this frequency falls off rapidly
with increasing neighborhood rank --- suggesting that optimization heuristics
could be improved by preferentially choosing links between very near
neighbors.  Our simulations show (see Figure \ref{fig_rl_neighbord2})
that for $d=2$ the decrease is astonishingly close to exponential.  We
may offer the following qualitative explanation for this behavior.  An
optimal tour will always try to use links to the closest neighbors possible.
While the constraint of a closed loop may force it in rare cases to use
neighbors of high rank, this will apply only to a very small 
number of links in the tour.  Connecting a point to, say, its $k$th-nearest
neighbor will for the most part be profitable only
when this neighbor is not much
further away than the $k-1$ nearer neighbors.  In other words, the lengths
from the point to its $k$ closest neighbors would have to be nearly degenerate.
Since a $k$-fold degeneracy of this sort is the product of $k-1$ unlikely
events, it is in fact quite natural that the probability of such an occurrence
is exponentially small in $k$.

We therefore conjecture that the neighborhood frequency function
will fall exponentially in $k$ at large $k$. We expect this 
behavior to hold 
in {\it any\/} dimension, and for that matter, in the Euclidean TSP
as well.  Similar and even stronger numerical results have been
reported \cite{Houdayer} in another link-based combinatorial
optimization problem, the matching problem.  An analytical calculation
of the neighborhood frequency may indeed turn out to be feasible using
the cavity approach, thus providing a theoretical prediction to accompany
our conjecture.  We consider this a significant open question.

\section{Numerical analysis: renormalized model}
\label{sec_numerrn}

In this section we will consider a different sort of random link TSP, 
proposed in
\cite{CBBMP}, allowing us to test numerically
the $1/d$ coefficient predicted by the cavity result
(\ref{eq_cavity}).
The approach involves introducing a mapping that shifts
and rescales all the lengths between cities.  By taking the limit
$d\to\infty$, one obtains a $d$-independent random link model
having an exponential distribution for its link lengths.
This ``renormalized'' model was outlined in \cite{CBBMP}; we present
it here in further detail.  We then perform a numerical study of the
model, which enables us to 
determine the large $d$ behavior of the standard $d$-dimensional
random link model.

Let us define $\langle D_1(N,d)\rangle$ to be the
distance between a city and its nearest neighbor, averaged over
all cities in the instance and over all instances in the
ensemble.\footnote{Note
that $\langle D_1(N,d)\rangle$ itself does not involve the notion of optimal
tours, or tours of any sort for that matter.}  For large $d$, it may be
shown \cite{CBBMP} that
\begin{equation}
\lim_{N\to\infty}N^{1/d}\langle D_1(N,d)\rangle =
\sqrt{\frac{d}{2\pi e}}\,(\pi d)^{1/2d} \left[ 1 - \frac{\gamma}{d} +
O\left(\frac{1}{d^2}\right)\right]
\label{eq_d1}
\end{equation}
where $\gamma$ is Euler's constant.  It is not surprising that
this quantity is reminiscent of (\ref{eq_rlexact}), since
$N^{1/d}\langle D_1(N,d)\rangle$ represents precisely a lower bound
on $\beta_{RL}(d)$.

In order to obtain the renormalized model,
consider a link length transformation making use of $\langle D_1(N,d)\rangle$.
For any instance with link lengths $l_{ij}$
(taken to have the usual distribution
(\ref{eq_dist})
corresponding to $d$ dimensions),
define new link lengths
$x_{ij}\equiv d [l_{ij} -\langle D_1(N,d)\rangle ] /
\langle D_1(N,d)\rangle$.
The $x_{ij}$ are ``lengths'' only in the
loosest sense, as they can be both positive and negative.  The optimal
tour in the $x_{ij}$ model will, however, follow the same ``path'' as
the optimal tour in the associated $l_{ij}$ model, since the transformation
is linear.  Its length $L_x(N,d)$ will simply be given in terms of
$L_{RL}(N,d)$ by:
\begin{eqnarray}
\label{eq_Lx_LRL}
L_x(N,d)&=&d\,\frac{L_{RL}(N,d)-N\langle D_1(N,d)\rangle}{\langle D_1(N,d)\rangle}
\mbox{, so}\\
L_{RL}(N,d)&=&N\langle D_1(N,d)\rangle\,\left[1+\frac{L_x(N,d)}{dN}\right]\mbox{.}
\end{eqnarray}
In the standard $d$-dimensional random link model,
$\beta_{RL}(d)=\lim_{N\to\infty} L_{RL}(N,d)/N^{1-1/d}$, so
\begin{eqnarray}
\beta_{RL}(d)&=&\lim_{N\to\infty}N^{1/d}\langle D_1(N,d)\rangle\,
\left[1+\frac{L_x(N,d)}{dN}\right]
\mbox{, and at large $d$, using (\ref{eq_d1}),}\nonumber \\
&=&\sqrt{\frac{d}{2\pi e}}\,(\pi d)^{1/2d} \left[ 1 - \frac{\gamma}{d} +
O\left(\frac{1}{d^2}\right)\right]\,\lim_{N\to\infty}\left[1+
\frac{L_x(N,d)}{dN}\right]\mbox{.}
\label{eq_rnbeta1}
\end{eqnarray}
As $\beta_{RL}(d)$ is a well-defined quantity, there must exist
a value $\mu(d)$ such that $\lim_{N\to\infty}L_x(N,d)/N=\mu(d)$.

Now, what will be the distribution of ``renormalized
lengths'' $\rho(x)$ corresponding to
$\rho(l)$?  From (\ref{eq_dist}) and the definition of the $x_{ij}$,
\begin{eqnarray}
\rho(x)&=& \frac{d\,\pi^{d/2}\,l^{d-1}}{\Gamma(d/2+1)}\,
\frac{\langle D_1(N,d)\rangle}{d}\mbox{, and substituting for $l$,}\nonumber \\
&=& \frac{\pi^{d/2}}{\Gamma(d/2+1)}\,\left(1+\frac{x}{d}\right)^{d-1}\,
\langle D_1(N,d)\rangle^d\mbox{.}
\end{eqnarray}
In the limit $N\to\infty$, we thus obtain from (\ref{eq_d1}) the large
$d$ expression:
\begin{eqnarray}
\rho(x)&\sim&\frac{\pi^{d/2}}{\Gamma(d/2+1)}\,\left(1+\frac{x}{d}\right)^{d-1}
N^{-1}\,\left(\frac{d}{2\pi e}\right)^{d/2}\sqrt{\pi d}\,
\left[1-\frac{\gamma}{d}+\cdots\right]^d\nonumber \\
&\sim&N^{-1}\left(1-\frac{\gamma}{d}\right)^d
\left(1+\frac{x}{d}\right)^{d-1}\left[1+O\left(\frac{1}{d}\right)\right]
\mbox{ by Stirling's formula}\nonumber \\
&\sim&N^{-1}e^{x-\gamma}\,\left[1+O\left(\frac{1}{d}\right)\right] .
\label{eq_rndist}
\end{eqnarray}
In the limit $d\to\infty$, $\rho(x)$ will be independent of $d$; the
same must then be true for $L_x(N,d)$, and consequently for $\mu(d)$.

Let us now {\it define\/} the renormalized model as being made up of
link ``lengths'' $x_{ij}$ in this limit.  This results in a somewhat
peculiar random link TSP, no longer containing the parameter $d$.  Its
link length distribution is given by the $d\to\infty$
limit of (\ref{eq_rndist}),
\begin{equation}
\label{eq_rndef1}
\rho(x)=N^{-1}\exp (x-\gamma),
\end{equation}
and its optimum tour length satisfies
\begin{equation}
\label{eq_rndef2}
\lim_{N\to\infty}\frac{L_x(N)}{N}=\mu,
\end{equation}
where we have dropped the $d$ argument from these (now $d$-independent)
quantities.  By performing direct simulations using the distribution
(\ref{eq_rndef1}) --- cut off beyond a threshold value of $x$,
as was done for $\rho_d(l)$ --- we may find the value of $\mu$ numerically.

\begin{figure}[!b]
\begin{center}
\epsfig{file=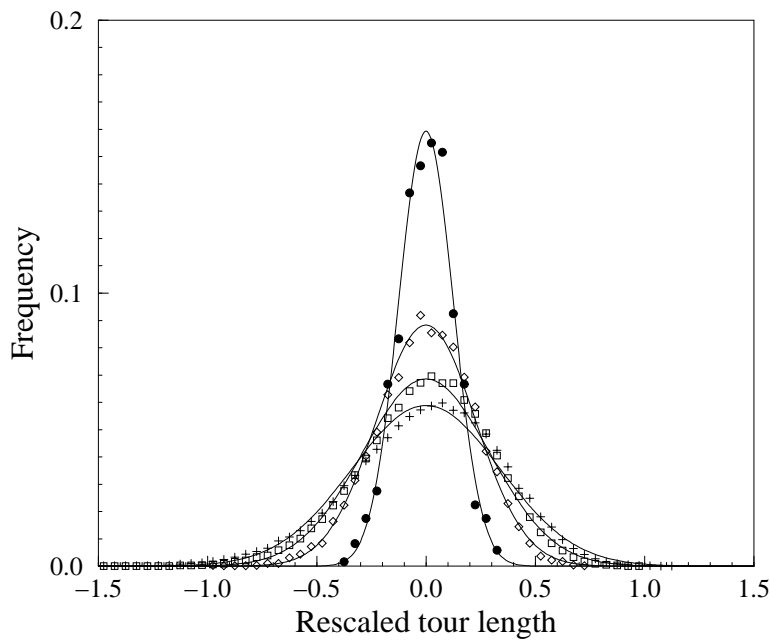,width=4in}
\caption{\small Distribution of renormalized random link rescaled tour length
$(L_{x}-\langle
L_{x}\rangle)/N$ for increasing values of N.  Plus signs show
$N=12$ (100,000 instances used), squares show $N=17$ (100,000 instances
used), diamonds show $N=30$ (4,000 instances used), and dots show $N=100$
(1,200 instances used).
Solid lines represent Gaussian fits for each value of $N$ plotted.}
\label{fig_rl_selfavrn}
\end{center}
\end{figure}

Finally, let us relate this renormalized model to the standard $d$-dimensional
random link model.  In light of (\ref{eq_rndef2}), we may rewrite
(\ref{eq_rnbeta1}) and obtain the result given in \cite{CBBMP}:
\begin{equation}
\beta_{RL}(d)=\sqrt{\frac{d}{2\pi e}}\,(\pi d)^{1/2d} \left[ 1 +
\frac{\mu - \gamma}{d} + O\left(\frac{1}{d^2}\right)\right]\mbox{.}
\label{eq_rnbeta2}
\end{equation}
The value of $\mu$ in the renormalized model therefore gives directly
the $1/d$ coefficient for the (non-renormalized) $\beta_{RL}(d)$.

\begin{figure}[t]
\begin{center}
\epsfig{file=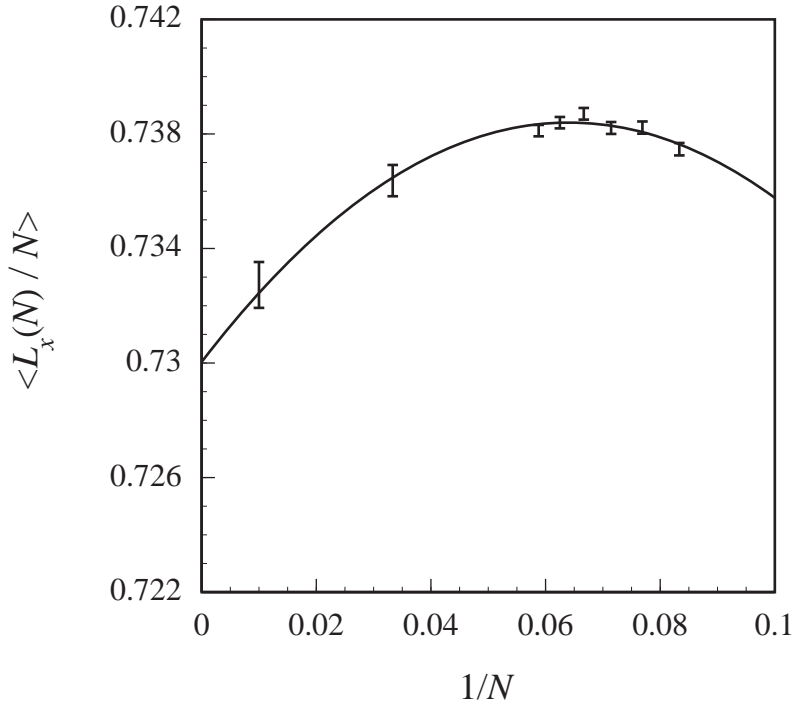}
\caption{\small Finite size scaling of renormalized model optimum.
Best fit ($\chi^2=5.23$) is given by: $\langle L_x(N)\rangle/N = 0.7300
(1 + 0.3575/N - 2.791/N^2)$.  Error bars show one standard deviation
(statistical error).}
\label{fig_rl_fssrn}
\end{center}
\end{figure}

We now carry out these direct simulations for the renormalized model.
Figures \ref{fig_rl_selfavrn} and \ref{fig_rl_fssrn} show our numerical
results.  In Figure \ref{fig_rl_selfavrn},
we see that just as in the $d=2$ case, the distribution of the optimum
tour length becomes sharply peaked at large $N$ and the asymptotic
limit $\mu$ is well-defined.  Via (\ref{eq_rnbeta2}), this provides
very good reason for believing that $\beta_{RL}(d)$ is well-defined
{\it for all\/} $d$, and that self-averaging holds for the random link
TSP in general.
In Figure \ref{fig_rl_fssrn}, we show the finite size
scaling of $\langle L_x(N)/N\rangle$.  The fit is again quite satisfactory
(with $\chi^2$=5.23 for 5 degrees of freedom), giving the asymptotic result
$\mu=0.7300\pm 0.0010$.  The resulting value for the $1/d$ coefficient
in $\beta_{RL}(d)$ is then $\mu-\gamma=0.1528\pm 0.0010$, in excellent
agreement (error under $0.3\%$) with the cavity prediction
$2-\ln 2-2\gamma = 0.1524\ldots$ given in (\ref{eq_cavity}).

\begin{figure}[t]
\begin{center}
\epsfig{file=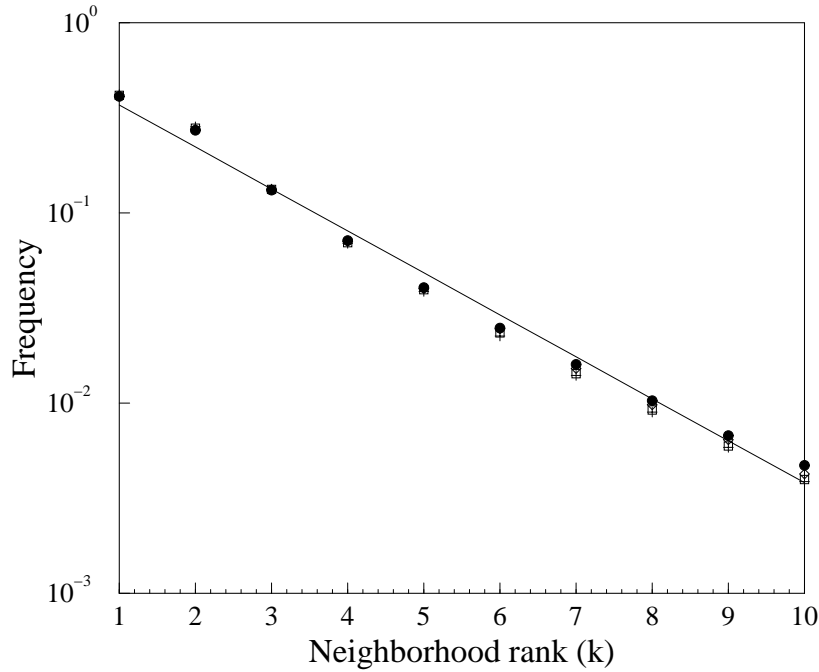,width=4in}
\caption{\small Frequencies with which $k$th-nearest neighbors are used in
optimal renormalized random link tours.  Plus signs show values for $N=12$,
squares for $N=17$, diamonds for $N=30$, and dots for $N=100$.  Best
exponential fit (straight line on log plot) is shown for $N=100$ data.}
\label{fig_rl_neighborrn}
\end{center}
\end{figure}

Again, as in the $d=2$ case, let us briefly consider the frequencies
of $k$th-nearest neighbors used in optimal tours.  
These frequencies are given in Figure \ref{fig_rl_neighborrn}
for the renormalized model.  Even though the exponential fit is
not as good as in the $d=2$ case, it is still striking here.
What does this tell us, in turn, about the standard random link TSP?
Recall that the renormalized
model arises from
the $d\to\infty$ limit of the $d$-dimensional (non-renormalized) model,
and that the
mapping (\ref{eq_Lx_LRL})
preserves the optimum tour for any given instance.
These $k$th-neighbor frequency results are thus
the $d\to\infty$ limiting frequencies for the $d$-dimensional random link
TSP (and most likely for the Euclidean TSP also). This gives further
support to our conjecture that the exponential law holds for all
$d$, and suggests as a consequence that the ``typical''
neighborhood rank $k$ used in optimal tours remains bounded for all $d$.

\section{Conclusion}
\label{sec_conclusion}

The random link TSP has interested theoreticians primarily because of its
analytical tractability, allowing presumably exact results that are not
possible in the more traditional Euclidean TSP.  Other than in the $d=1$ case,
however, it has attracted little attention.
In this paper we have provided a numerical study of the random link TSP
that was lacking up to this point, addressing important unanswered
questions.  Through simulations, we have tested the validity of the
theoretical predictions derived using the cavity method.
While in other disordered systems, such as spin glasses, the replica symmetric
solution gives values of macroscopic quantities that are inexact (typically
by several percent), in the random link TSP it shows all signs of
being exact.  
We have studied various link-based quantities at $d=2$ and found
that the numerical results confirm the cavity predictions to within
$0.1\%$.  Furthermore,
we have confirmed, by way of simulations on a renormalized random link model,
that the analytical cavity solution gives a large $d$ expansion for the
optimum tour length whose $1/d$ coefficient is correct to within well
under $1\%$.  The excellent agreement found at $d=1$
\cite{KrauthMezard,Sourlas},
$d=2$, and to $O(1/d)$ at large $d$, then suggest strongly
that the cavity predictions are exact.  This provides indirect
evidence that the assumption of replica symmetry --- on which the
cavity calculation is based --- is indeed justified for the TSP.

Finally, our random link simulations have pointed to a surprising
numerical result.  If one considers the links in optimal tours as links
between $k$th-nearest neighbors, at $d=2$ the frequency with which the
tour uses neighborhoods of rank $k$ decreases with $k$ as almost a perfect
exponential.
Encouraged by similar results in the renormalized model, we conjecture
that this property holds true for all $d$, as well as in the Euclidean
TSP.  As no theoretical calculation presently explains the phenomenon, we
would welcome further investigation along these lines.

\section*{Acknowledgments}
\label{sect_ack}
Thanks go to J.~Houdayer and N.~Sourlas for their insights and suggestions
concerning $k$th-nearest neighbor statistics, and to J.~Boutet de Monvel
for his many helpful remarks on the cavity method.  AGP acknowledges
the hospitality of the Division de Physique Th\'eorique, Institut de Physique
Nucl\'eaire, Orsay, where much of this work was carried out.
OCM acknowledges support from the Institut Universitaire de France.
The Division de Physique Th\'eorique is an {\it Unit\'e
de Recherche des Universit\'es Paris XI et Paris VI associ\'ee au CNRS\/}.

\end{document}